\documentclass[5p, times, twocolumn, number, sort&compress]{elsarticle-accepted}

\usepackage[T1]{fontenc}
\usepackage[colorlinks=true]{hyperref}
\usepackage{graphics}
\usepackage{latexsym}
\usepackage{amssymb}
\usepackage{amsfonts}
\usepackage{amsmath}
\usepackage{amsthm}
\usepackage{multirow}
\usepackage{longtable}
\usepackage{mathtools}
\usepackage{braket}
\usepackage{siunitx}
\usepackage{xcolor}
\usepackage{nicefrac}
\usepackage{orcidlink}

\newcommand{\etal}{\textit{et al.}}

\journal{Physics Letters B}

\begin{document}

\title{Electromagnetic moments of the odd-mass nickel isotopes $^{59-67}$Ni}

\author[1]{P. Müller\orcidlink{0000-0002-4050-1366}}
\author[1]{S. Kaufmann\orcidlink{0009-0006-3149-6434}}
\author[1,2,3]{T. Miyagi\orcidlink{0000-0002-6529-4164}}
\author[4]{J. Billowes}
\author[4]{M. L. Bissell}
\author[2]{K. Blaum\orcidlink{0000-0003-4468-9316}}
\author[5]{B. Cheal\orcidlink{0000-0002-1490-6263}}
\author[6]{R.~F.~Garcia~Ruiz\fnref{MIT}}
\author[7]{W. Gins}
\author[1]{C. Gorges}
\author[2,6]{H.~Heylen\orcidlink{0009-0001-7278-2115}}
\author[7]{A. Kanellakopoulos\orcidlink{0000-0002-6096-6304}}
\author[6]{S. Malbrunot-Ettenauer}
\author[2,8]{R. Neugart}
\author[6,7]{G. Neyens\orcidlink{0000-0001-8613-1455}}
\author[1]{W. N\"ortersh\"auser\orcidlink{0000-0001-7432-3687}\corref{cor1}}
\ead{wnoertershaeuser@ikp.tu-darmstadt.de}
\author[1]{T. Ratajczyk}
\author[2,6,11]{L. V. Rodr\'iguez\orcidlink{0000-0002-0093-2110}}
\author[9]{R.~S\'anchez\orcidlink{0000-0002-4892-4056}}
\author[10]{S. Sailer}
\author[1,2,3]{A. Schwenk\orcidlink{0000-0001-8027-4076}}
\author[8]{L. Wehner}
\author[5]{C. Wraith}
\author[4]{L. Xie}
\author[7]{Z. Y. Xu}
\author[7,12]{X.~F.~Yang\orcidlink{0000-0002-1633-4000}}
\author[11]{D. T. Yordanov}

\cortext[cor1]{Corresponding author}
\fntext[MIT]{present address: Massachusetts Institute of Technology, MA 02139, Cambridge, USA}
\affiliation[1]{organization={Institut f\"ur Kernphysik, Technische Universit\"at Darmstadt},
    postcode={D-64289},
    city={Darmstadt},
    country={Germany}}
\affiliation[2]{organization={Max-Planck-Institut f\"ur Kernphysik},
    postcode={D-69117},
    city={Heidelberg},
    country={Germany}}
\affiliation[3]{organization={ExtreMe Matter Institute EMMI, GSI Helmholtzzentrum f\"ur Schwerionenforschung GmbH},
    postcode={D-64291},
    city={Darmstadt},
    country={Germany}}
\affiliation[4]{organization={School of Physics and Astronomy, The University of Manchester},
    postcode={M13 9PL},
    city={Manchester},
    country={United Kingdom}}
\affiliation[5]{organization={Oliver Lodge Laboratory, Oxford Street, University of Liverpool},
    postcode={L69 7ZE},
    city={Liverpool},
    country={United Kingdom}}
\affiliation[6]{organization={Experimental Physics Department, CERN},
    postcode={CH-1211},
    city={Geneva 23},
    country={Switzerland}}
\affiliation[7]{organization={KU Leuven, Instituut voor Kern- en Stralingsfysica},
    postcode={B-3001},
    city={Leuven},
    country={Belgium}}
\affiliation[8]{organization={Institut f\"ur Kernchemie, Johannes Gutenberg-Universit\"at Mainz},
    postcode={D-55128},
    city={Mainz},
    country={Germany}}
\affiliation[9]{organization={GSI Helmholtzzentrum f\"ur Schwerionenforschung GmbH},
    postcode={D-64291},
    city={Darmstadt},
    country={Germany}}
\affiliation[10]{organization={Technische Universit\"at M\"unchen},
    postcode={D-80333},
    city={M\"unchen},
    country={Germany}}
\affiliation[11]{organization={Institut de Physique Nucl\'eaire, CNRS-IN2P3, Universit\'e Paris-Sud, Universit\'e Paris-Saclay},
    postcode={91406},
    city={Orsay},
    country={France}}
\affiliation[12]{organization={School of Physics and State Key Laboratory of Nuclear Physics and Technology, Peking University},
postcode={100871},
city={Beijing},
country={China}}

\doi{10.1016/j.physletb.2024.138737}
\license{http://creativecommons.org/licenses/by/4.0/}

\begin{abstract}
The magnetic dipole and the spectroscopic quadrupole moments of the nuclear ground states in the odd-mass nickel isotopes $^{59-67}$Ni have been determined using collinear laser spectroscopy at the CERN-ISOLDE facility. They are compared to \textit{ab initio} valence-space in-medium similarity renormalization group (VS-IMSRG) calculations including contributions of two-body currents as well as to shell-model calculations. The two-body-current contributions significantly improve the agreement with experimental data, reducing the mean-square deviation from the experimental moments by a factor of 3 to 5, depending on the employed interaction. For all interactions, the largest contributions are obtained for the $\nicefrac{5}{2}^-$ ($\nicefrac{7}{2}^-$) isotopes $^{65}$Ni ($^{55}$Ni), which is ascribed to the high angular momentum of the $f$ orbitals. Our results demonstrate that the inclusion of two-body-current contributions to the magnetic moment in an isotopic chain of complex nuclei can be handled by the VS-IMSRG and can outperform phenomenological shell-model calculations using effective $g$-factors in the nickel region.
\end{abstract}

\begin{keyword}
Collinear laser spectroscopy \sep Electromagnetic moments \sep Nickel isotopes \sep Ab initio calculation \sep Valence-space in-medium similarity renormalization group \sep Nuclear shell model
\end{keyword}

\maketitle

\section{Introduction}

An accurate description of nuclei is essential for understanding the formation of matter in our universe. Major astrophysical processes such as core-collapse supernovae and neutron star mergers depend on the states and dynamics of nucleonic matter~\cite{Schatz.2022}. To evolve our knowledge on nuclear structure applied in the analysis of astrophysical observations, benchmark values from laboratory experiments are required.

Nickel ($Z=28$) lies in a particularly interesting region of the nuclear chart from several aspects. From an astrophysics point of view, nickel is at the border between those elements created by burning in stars and the heavier ones created in explosive events \cite{Schatz.2022}. Nickel has a magic proton number and its isotopic chain spans over three ``conventional'' magic neutron numbers ($N=20, 28, 50$). The doubly magic $^{56}$Ni plays a major role in stellar nucleosynthesis \cite{Truran.1967, Timmes.2003, Chieffi.2017, Thielemann.2018}. Recently, the nuclear charge radii of the neutron-deficient isotopes $^{54-56}$Ni \cite{Sommer.2022} and the neutron-rich isotopes $^{58-68,70}$Ni \cite{Malbrunot-Ettenauer.2022} were determined using collinear laser spectroscopy at NSCL/MSU and ISOLDE/CERN, respectively. Density functional, coupled-cluster and valence-space in-medium similarity renormalization group (VS-IMSRG) calculations were performed and used to benchmark different Hamiltonians and many-body methods. From constraining the properties of nuclear matter in neutron stars using $^{54}$Ni \cite{Pineda.2021} to investigating the correlation between the charge radius of $^{68}$Ni and the nuclear dipole polarizability \cite{Kaufmann.2020}, valuable information about nuclear interactions that also govern astrophysical processes can be gained \cite{Schatz.2022}.

Similar to charge radii, electromagnetic moments, which are obtained from optical spectra with high accuracy, also provide important information about the nuclear state \cite{Neugart.1981, Otten.1989, Neyens.2003, Campbell.2016}. The nuclear magnetic moment $\mu$, often represented by the $g$-factor $g=|\mu|/I\mu_N$, is an excellent probe for the single-particle structure of the nucleus \cite{Neyens.2003}. Close to a magic number, magnetic dipole moments are expected to be relatively close to the single-particle expectation value, the so-called Schmidt value. Experimental nuclear moments of even-odd nuclei are in practically all cases smaller (in magnitude) than the Schmidt values \cite{Schmidt.1937}, even for systems close to doubly magic nuclei \cite{Andl.1982, Ohtsubo.1996, Yordanov.2020}. This is partially ascribed to configuration mixing, which usually becomes stronger when moving away from shell-closures since particle-hole excitations are less costly in terms of energy. Deviations observed in the direct neighborhood of doubly magic nuclei (doubly magic $\pm 1$) are ascribed to an influence of the medium on the proton's (or neutron's) magnetic moment and are accounted for by the ad hoc introduction of an effective $g_\mathrm{eff}$ factor in shell-model calculations \cite{Castel.1990, Brown.2001, Neyens.2003}. $g_\mathrm{eff}$ includes the influence of two-body currents (2BC) as well as first- and second-order core polarization effects \cite{Castel.1990}. For even-odd nuclei in the nickel region, a value of $g_\mathrm{eff} = 0.7 g_\mathrm{free}$ is usually used \cite{Vingerhoets.2010, Kanellakopoulos.2020}.

Recently, we have reported on nuclear moments of Sb isotopes ($Z=51$) and compared them with results from valence-space in-medium similarity renormalization group (VS-IMSRG) calculations \cite{Lechner.2023}. While the agreement was imperfect, the trend was reasonably reproduced. In addition, it was demonstrated that the phenomenological shell-model and VS-IMSRG wave function parts relevant to the magnetic moments are similar. Therefore, the conclusions were that an improvement in the magnetic moment operator is required, and the most probable reason is due to missing 2BC contributions. In the weak sector, it was demonstrated that 2BCs need to be included to reproduce the measured Gamow-Teller transition strength~\cite{Gysbers2019}. Similarly, for light nuclei, it was shown that electromagnetic 2BCs improve the agreement with measured magnetic dipole observables~\cite{Pastore2013,Friman-Gayer2021}. Most recently, it was reported that 2BC effects are also significant in medium-mass and heavy nuclei~\cite{Acharya2023, Miyagi2023}. Therefore, we expect a similar improvement from 2BC effects over the nickel isotopic chain.

Here, we present new and improved experimental magnetic dipole moments in the odd-mass nickel isotopes $^{59,63,65,67}$Ni as well as first measurements of the spectroscopic quadrupole moments of $^{59,65}$Ni. The values are obtained from the determined hyperfine structure constants using the known $^{61}$Ni moments, which were critically reevaluated. We compare these experimental results to VS-IMSRG calculations including 2BC contributions as well as to phenomenological shell-model results.

\section{Experiment}
\label{sec:experiment}

A detailed description of the general experimental setup of the COLLAPS experiment at ISOLDE/CERN can be found in~\cite{Neugart.2017}. Results on isotope shifts from the two nickel beamtimes in 2016 and 2017 were published previously \cite{Kaufmann.2020, Malbrunot-Ettenauer.2022}. Here, we provide some more details of the experimental procedure, the differences between both beamtimes, and the lineshape analysis.
Nickel isotopes were produced by 1.4-GeV proton pulses impinging on a uranium carbide target. The physical and chemical properties of Ni lead to release times that are long compared to the recurrence time of the proton bunches. This allowed to gate for a delay after each proton trigger to release unwanted Ga from the target. Nickel atoms effusing into the ion source were ionized using in-source resonant laser ionization with RILIS \cite{Marsh.2014} and then accelerated to $30$\,keV (2016) and $40$\,keV (2017) of kinetic energy. Subsequently, the ions were mass-separated using the high-resolution mass separator (HRS) as well as cooled and accumulated for $10-100$\,ms with the radio-frequency quadrupole cooler-buncher ISCOOL \cite{Franberg.2008}. 
Ion bunches of typically $5\,\mu$s-duration were sent to the COLLAPS beamline where they were superimposed with a copropagating laser beam and neutralized in a charge-exchange cell (CEC) \cite{Mueller.1983, Klose.2012} filled with sodium (2016) or potassium vapor (2017). Neutral Ni atoms that reside in the $3d^9 4s\,^3$D$_3$ meta-stable state after the CEC \cite{Ryder.2015} are resonantly excited into the $3d^94p\,^3$P$_2$ state at a laboratory wavelength of about 352.199\,nm (352.143\,nm) at 30\,keV (40\,keV) beam energy, respectively. The excited state predominantly decays back into the $^3$D$_3$ level by spontaneous emission of a photon. These signal photons were detected using four photo-multiplier tubes situated directly behind the CEC \cite{Kreim.2014}. Detection is time-gated to the ion-bunches in order to suppress stray light from the laser beam. Laser light was generated with a frequency-doubled single-mode continuous-wave Ti:sapphire laser stabilized to a high-precision wavelength meter (High Finesse, WSU-10) \cite{Verlinde.2020,Konig.2020} that was regularly calibrated with a frequency-stabilized helium-neon laser (2016) or a diode laser locked to a Rb line (2017).
In the first beamtime, COLLAPS was controlled by the MCP data acquisition (DAQ) system and a new DAQ (\texttt{TILDA}) was run parasitically in parallel. Hence, the time gate for photon detection had to be pre-set in hardware to about 10\,\textmu{}s. The variable delay for the isotope's time-of-flight (TOF) from ISCOOL to the detection region had to be adjusted for each isotope individually. \texttt{TILDA} allowed monitoring the TOF to ensure that the buncher was not overfilling. In the second beamtime, \texttt{TILDA} became the main DAQ and fully controlled all devices. Here, ranges of interest for the photon detection are optimized by software gates in the post-beamtime analysis. Thus, the background suppression by bunched beam operation is exploited to its full potential and the gates for photon detection were reduced from 10\,\textmu s to 3\,\textmu s. For details on \texttt{TILDA} see~\cite{Kaufmann.2019,Kanellakopoulos.2020,Malbrunot-Ettenauer.2022}. All isotopes were measured in both beamtimes with the exceptions of $^{59,63}$Ni, which were only addressed in 2016.

\begin{figure}[!t]
\centering
\includegraphics[width=\linewidth]{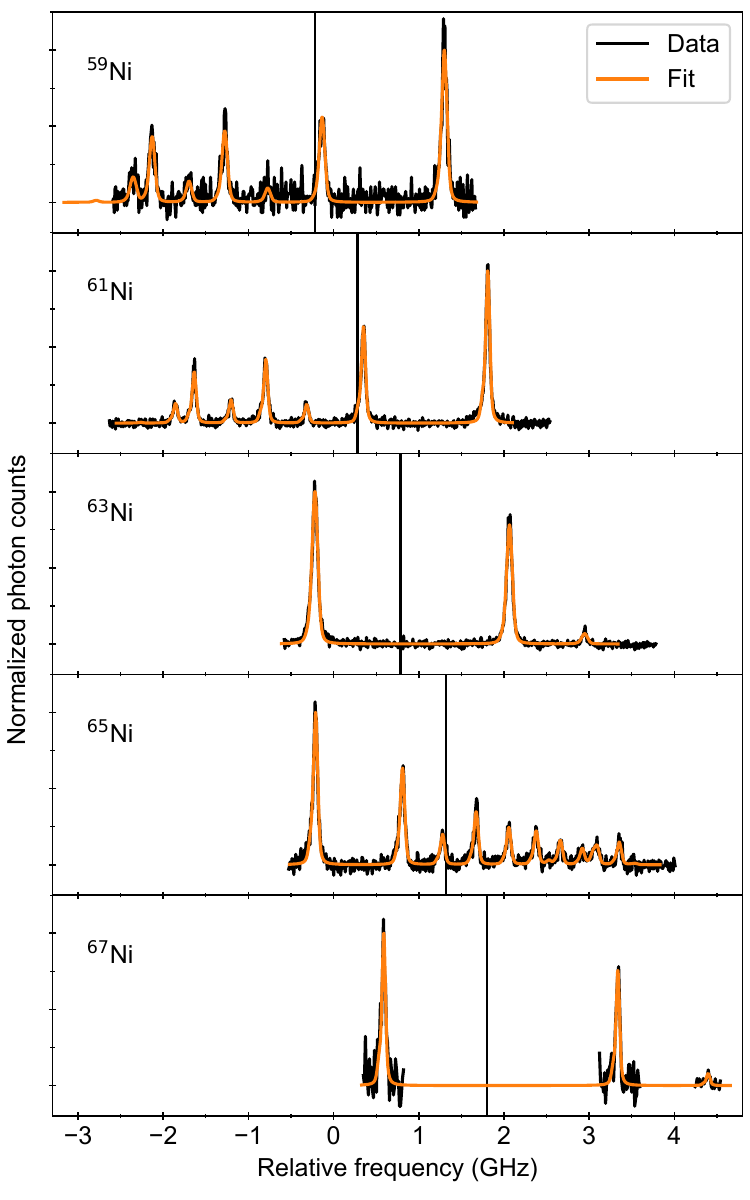}
\caption{\label{fig:spectra} Hyperfine spectra of the $3d^9 4s\,^{3}\mathrm{D}_3 \rightarrow 3d^9 4p\,^3\mathrm{P}_2$ transition in the odd-mass nickel isotopes $^{59-67}$Ni. The black lines mark the center of gravity of the hyperfine spectra and correspond to the isotope shifts relative to $^{60}$Ni which can be found in \cite{Malbrunot-Ettenauer.2022}. A sum of Voigt profiles, each with a small satellite peak at a fixed distance to the main peak, was fitted to the data.}
\end{figure}

\section{Analysis}
\label{sec:results}

\subsection{Spectral Lineshape}

\begin{figure}[t]
\centering
\includegraphics[width=\linewidth]{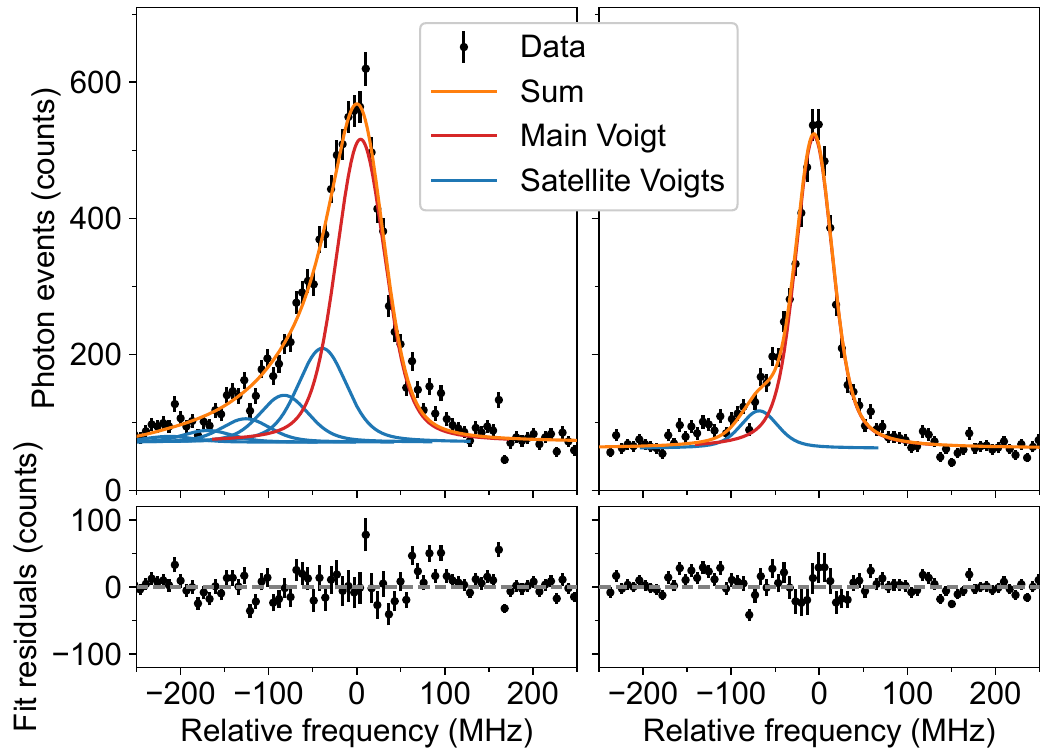}
\caption{\label{fig:ref-spectrum} Example spectra of the reference isotope $^{60}$Ni taken under optimal conditions (right) and with a higher pressure in the charge exchange cell due to elevated temperatures as it happened at the beginning of the 2017 beamtime (left). In the latter case, a sequence of five sidepeaks was used to properly describe the resonance signal. Optimal values for the peak offsets and intensity ratios were determined from the spectra of the reference isotope and applied for all other isotopes. For details see text.}
\end{figure}

The hyperfine spectra of the $3d^9 4s\,^{3}\mathrm{D}_3 \rightarrow 3d^9 4p\,^3\mathrm{P}_2$ transition in the odd-mass nickel isotopes $^{59,61,63,65,67}$Ni were taken to determine the magnetic dipole and spectroscopic electric quadrupole moments as well as the nuclear spin of $^{67}$Ni. Spectra of all measured odd-mass isotopes are shown in Fig.\,\ref{fig:spectra}. 
Resonance lines were fitted with the \texttt{Python}-based \texttt{PolliFit} framework which was developed at TU Darmstadt and uses the \texttt{Python} package \texttt{scipy} for fitting a function using least-squares minimization \cite{scipy}. The experimental data exhibits a small asymmetry with a tail at lower energies as it is usual for collinear laser spectroscopy after a charge-exchange cell. This is explained by elastic and inelastic collisions inside the CEC that always cause a reduction in kinetic energy \cite{Bendali.1986,Krieger.2016}. To accommodate for this asymmetry a satellite-Voigt is added to the main Voigt profile to reproduce the shoulder and to get an improved $\chi^2$. Note that despite the effort to define a consistent lineshape, which is described in the following, a simple symmetric Voigt profile could be used that would consistently yield the same isotope shifts and hyperfine structure constants. This is a consequence of fixing the asymmetry for the isotope and its reference. However, eliminating systematic structures in the fit residuals has the advantage to yield more thorough statistical uncertainties as they are scaled by the reduced $\chi^2$.

Since the offset between the main peak and the satellite is associated with an energy reduction, it is fitted in energy and not in frequency space. In this way, the offset is to first order equal for all isotopes, while it depends in frequency space on their individual differential Doppler-shifts. This energy offset was found by first fitting all spectra of the reference isotope with this offset as one of the optimization parameters, yielding a mean value for the offset of $-5.175(5)$\,eV and $-4.76(5)$\,eV for the 2016 and the 2017 beamtime, respectively. This offset was then fixed for all spectra, including the reference isotopes, while the intensity of the satellite-Voigt was left free for optimization. An example spectrum of the reference isotope is depicted in the right panel of Fig.\,\ref{fig:ref-spectrum}.
In the first part of the 2017 beamtime, while measuring $^{61}$Ni, the CEC was overheated and several collision processes occurred while traveling through the gas. Thus, a stronger asymmetry was observed that could not be described by a single satellite anymore. Instead, a total of 5 equidistant satellites having the same width and decreasing relative intensity $I_{n+1}=\nicefrac{1}{2}I_n={\left(\nicefrac{1}{2}\right)}^{n-1}I_1$, where $I_1$ is the intensity of the first satellite, was used. For the fitting, all satellite-Voigts were chosen to exhibit the same widths as the main peak and their distances were chosen to be equal to $-3.32(5)$\,eV. An example profile of the reference isotope $^{60}$Ni is shown in the left panel of Fig.\,\ref{fig:ref-spectrum}.

The Voigt profiles have a mean full width at half maximum (FWHM) of $67(12)\,$MHz with a Gaussian FWHM of $55(12)\,$MHz and a Lorentzian FWHM of $21.14\,$MHz. The latter was fixed for better convergence. The natural linewidth, determined by adding the Einstein $A_{ki}$ coefficients of all transitions with $3d^9 4p\,^3\mathrm{P}_2$ as the upper state, is assumed to be $17.4\,$MHz \cite{NIST}.
Peak intensities were treated as free fit parameters wherever statistics allowed it. Peaks that lie outside of the data range or vanish in the noise had their relative intensities scaled to the respective theoretical transition strengths. The fitted peak intensities only show slight deviations from the theoretical values such that optical pumping effects are minimal and do not prevent unambiguous nuclear spin assignments. Possible shifts due to quantum interference effects between close lying excited states are smaller than $0.1\,$MHz for all hyperfine constants and thus negligible. The given FWHM and statistics result in a minimal statistical uncertainty of $0.4\,$MHz per spectrum for peak position parameters, which is close to the usual limit for the uncertainty-to-FWHM ratio in CLS.

The magnetic-dipole parameters $A_l\equiv A(^3$D$_3)$, $A_u\equiv A(^3$P$_2)$ and the electric-quadrupole parameters $B_l\equiv B(^3$D$_3)$, $B_u\equiv B(^3$P$_2)$ of the lower and the upper state of the transition, respectively, were obtained from the individual fits and averaged weighted by their fitting uncertainty. With the exception of $^{65}$Ni we found that $0.1 \leq \chi_\mathrm{red}^2 \leq 1$. For $^{65}$Ni ($\chi_\mathrm{red}^2=1.35$), the uncertainty was rescaled by $\sqrt{\chi_\mathrm{red}^2}$. The $A$ and $B$ factors determined in that manner for both beamtimes agreed well within their respective uncertainties. For $^{61}$Ni, the final values obtained for $A_l$ are $454.8(4)$ and $455.0(3)$\,MHz and for $A_u$ the numerical value with $177.1(4)$\,MHz was indeed identical in both beamtimes, indicating that the chosen line profiles were reasonable to extract peak distances. The final values for $A_l$, $A_u$ and $B_l$, $B_u$ were obtained as the weighted average of the two values and the smaller uncertainty was taken as the final uncertainty. Results are listed in Tab.\,\ref{tab:hfs_parameters}. We note that the isotope shifts determined from the same fits and published earlier \cite{Malbrunot-Ettenauer.2022} were combined similarly and agreed nicely with those of independent measurements at NSCL/MSU~\cite{Konig.2021b}.      

Within uncertainties all ratios agree well with the average of $A_u/A_l = 0.3894(6)$, showing no sign of higher order nuclear contributions to the $A$ constants beyond the point-like nuclear magnetic dipole moment, enabling the use of Eq.\,\eqref{eq:m}. The ratio is also in accordance with the value of $0.389(17)$ measured for $^{55}$Ni at NSCL/MSU \cite{Sommer.2022}. The $A$ and $B$ factors of the lower electronic level are in excellent agreement with existing literature values for the stable isotopes, particularly with the very precise $A_l$ and $B_l$ of $^{61}$Ni as determined from atomic-beam magnetic resonance spectroscopy \cite{Childs.1968}. The lower level also has a higher sensitivity to the magnetic dipole and the electric quadrupole moment, which is obvious by the larger magnitude of the respective hyperfine parameters. For these reasons, $A_l$ and $B_l$ will dominate the determination of the electromagnetic moments of all other odd-mass Ni isotopes. The nuclear spin of $^{67}$Ni, is specified in round brackets in \cite{Stone.2019}, which indicates that it is not sufficiently confirmed. From our optical spectrum of $^{67}$Ni, we can unambiguously determine its spin as $\nicefrac{1}{2}$ since any other spin contradicts the given peak intensities and splittings.

\begin{table}[t]
\centering
\caption{\label{tab:hfs_parameters} Hyperfine structure parameters determined from the fits to the laser spectroscopy data as shown in Fig.\,\ref{fig:spectra}. All values are weighted means of the individual fit results. In case of $\chi_\mathrm{red}^2>1$, the uncertainty of the weighted mean was scaled by $\sqrt{\chi_\mathrm{red}^2}$. All values are given in MHz.}
\resizebox{\columnwidth}{!}{%
\begin{tabular}{r r r r r r}
\multicolumn{1}{c}{$A$} & \multicolumn{1}{c}{$I^\pi$} & \multicolumn{1}{c}{$A_l$} & \multicolumn{1}{c}{$B_l$} & \multicolumn{1}{c}{$A_u$} & \multicolumn{1}{c}{$B_u$} \\
\hline
59 & $\nicefrac{3}{2}^-$ & $-452.7(1.1)$ & $-56.7(6.8)$ & $-176.1(1.6)$ & $-31.5(5.5)$ \\
61 & $\nicefrac{3}{2}^-$ & $-454.9(0.3)$ & $-102.6(1.7)$ & $-177.1(0.4)$ & $-50.6(1.6)$ \\
&  & $-454.972(3)^\text{a}$ & $-102.951(16)^\text{a}$ & \multicolumn{1}{c}{--}  & \multicolumn{1}{c}{--} \\
63 & $\nicefrac{1}{2}^-$ & $904.5(1.1)$ & \multicolumn{1}{c}{--} & $352.3(1.6)$ & \multicolumn{1}{c}{--} \\
&  & $903(11)^\text{b}$ & \multicolumn{1}{c}{--} & \multicolumn{1}{c}{--}  & \multicolumn{1}{c}{--} \\
65 & $\nicefrac{5}{2}^-$ & $276.6(0.3)$ & $-59.4(3.4)$ & $107.8(0.3)$ & $-28.7(2.6)$ \\
67 & $\nicefrac{1}{2}^-$ & $1088.6(0.6)$ & \multicolumn{1}{c}{--} & $423.5(0.2)$ & \multicolumn{1}{c}{--}
\end{tabular}}
\flushleft
{\footnotesize $^\text{a}$ taken from \cite{Childs.1968}\\
\footnotesize $^\text{b}$ taken from \cite{Dyachkov.2017}}
\end{table}

\subsection{Determination of the Nuclear Moments}

The magnetic dipole and spectroscopic quadrupole moments $\mu, Q$ of the nickel isotopes can be calculated from the measured hyperfine structure parameters if the respective moments $\mu_\mathrm{ref}, Q_\mathrm{ref}$ of a reference isotope are available. Under the assumption that the contribution of the spatial extent of the nucleus on the hyperfine constants is negligible, which is sufficiently fulfilled at our level of accuracy, as evidenced by the constant $A$ and $B$ ratios between the upper and the lower state across all isotopes, the required relations are

\begin{align}
\label{eq:m}
\mu &= \mu_\mathrm{ref}\frac{A}{A_\mathrm{ref}}\frac{I}{I_\mathrm{ref}}\\
\label{eq:q}
Q &= Q_\mathrm{ref}\frac{B}{B_\mathrm{ref}},
\end{align}

with $A, B, A_\mathrm{ref}, B_\mathrm{ref}$ the hyperfine-structure constants and $I, I_\mathrm{ref}$  the nuclear spins of the isotope and the reference isotope, respectively. Here, $^{61}$Ni was chosen as the reference isotope and the high-precision $A$ and $B$ factors of the lower state determined in \cite{Childs.1968} were used as reference values. The last ingredient that is needed are the nuclear moments of $^{61}$Ni. The most precise nuclear magnetic dipole moments of stable nuclei are usually obtained using nuclear magnetic resonance measurements since they measure nuclear precession frequencies in external magnetic fields. However, the accuracy of the extracted nuclear moment does also depend on the knowledge of the magnetic field at the nuclear site. The external field is modified by atomic effects, particularly the diamagnetic shielding of the external field by the electron shell of the atom under investigation, as well as by other atoms in the chemical environment. This diamagnetic shielding has been treated in an approximate way already in the early tabulations by Fuller \etal\ \cite{Fuller.1969, Fuller.1976}, where corrections obtained by Dickinson \cite{Dickinson.1950} were applied and an uncertainty of 5\% was assumed. The table by Raghavan \cite{Raghavan.1989} used updated corrections from \cite{Johnson.1983} and treated these without uncertainties. This lead to magnetic moments that appeared to be very accurate and the same procedure was adapted in the later tabulations by Stone \cite{Stone.2005,Stone.2014}. However, several results obtained in more recent NMR measurements, e.g., \cite{Antusek.2005,Skripnikov.2018,Fella.2020}, indicated discrepancies being much larger than the stated uncertainties. 
Only in the latest version of the INDC table~\cite{Stone.2019}, the uncertainty of the diamagnetic correction has been discussed more thoroughly. For 29 cases, where modern calculations were available the new corrections with larger uncertainties have been used, while for the others (including $^{61}$Ni), a scaling of the previous shielding factors to 75\% of the ``historical'' values was used and a general uncertainty of 10\% was introduced. For Ni, this resulted in an incremental increase of the uncertainty while the new magnetic moment of $-0.74965(5)\,\mu_\mathrm{N}$ \cite{Stone.2019} in $^{61}$Ni is $7\sigma$ smaller than the previously tabulated one of $-0.75002(4)$~\cite{Raghavan.1989,Stone.2014}.

\begingroup
\begin{table*}[t]
\centering
\caption{\label{tab:m} Magnetic dipole moments $\mu$ of the odd-mass nickel isotopes $^{59-67}$Ni determined using Eq.\,\eqref{eq:m} with $^{61}$Ni as the reference isotope. Results of shell-model calculations using the \texttt{NuShellX@MSU} \cite{Brown.2014}, and \texttt{KSHELL} \cite{KSHELL} codes (see text) and three phenomenological interactions tailored for the $f_{\nicefrac{7}{2}}$, $f_{\nicefrac{5}{2}}$, $p_{\nicefrac{3}{2}}$, $p_{\nicefrac{1}{2}}$, $g_{\nicefrac{9}{2}}$ region with different active spaces are given. Moreover, \textit{ab initio} VS-IMSRG results based on three different NN+3N interactions, including both one- and two-body-current contributions, are presented. All values, except those in the last line, are given in units of the nuclear magneton $\mu_{\mathrm{N}}$.\\}
\resizebox{\textwidth}{!}{%
\begin{tabular}{l l l l l r r r r r r}
$A$ & $I^\pi$ & This work & Literature & References & \multicolumn{3}{c}{Shell Model}  & \multicolumn{3}{c}{VS-IMSRG} \\
&  &  &  & & \multicolumn{1}{c}{JUN45} & \multicolumn{1}{c}{jj44b} & \multicolumn{1}{c}{GXPF1A} & \multicolumn{1}{c}{1.8/2.0 (EM)} & \multicolumn{1}{c}{$\Delta$N$^2$LO$_\mathrm{GO}$} & \multicolumn{1}{c}{N$^3$LO$_\mathrm{lnl}$} \\
\hline
55 & $\nicefrac{7}{2}^-$ & $-1.105(20)^\text{a}$ & $-1.108(20)$ & \cite{Sommer.2022} & \multicolumn{1}{c}{--} & \multicolumn{1}{c}{--} & $-0.994$ & $-0.760$ & $-0.766$ & $-0.654$ \\
57 & $\nicefrac{3}{2}^-$  & \multicolumn{1}{c}{--}  & $-0.7975(14)$ & \cite{Ohtsubo.1996} & $-1.339$ & $-1.339$ & $-0.776$ & $-0.844$ & $-0.813$ & $-0.590$\\
59 & $\nicefrac{3}{2}^-$ & $-0.7439(15)$ & \multicolumn{1}{c}{--} & & $-0.873$ & $-1.226$ & $-0.670$ & $-0.916$ & $-0.793$ & $-0.675$ \\
61 & $\nicefrac{3}{2}^-$ & $-0.7472(6)$ & $-0.7473(4)$ & \cite{Antusek.2020} & $-0.916$ & $-1.145$ & $-0.493$ & $-0.955$ & $-0.905$ & $-0.723$ \\
63 & $\nicefrac{1}{2}^-$ & $0.4950(4)$ & $0.496(5)$ & \cite{Dyachkov.2017} & $0.446$ & $0.458$ & $0.508$ & $0.591$ & $0.523$ & $0.658$ \\
65 & $\nicefrac{5}{2}^-$ & $0.7575(7)$ & $0.69(6)$ & \cite{Krane.1976, Stone.2014} & $0.858$ & $0.809$ & $0.786$ & $0.826$ & $0.749$ &  $0.795$\\
67 & $\nicefrac{1}{2}^-$ & $0.5961(7)$ & $0.601(5)$ & \cite{Rikovska.2000, Stone.2014} & $0.452$ & $0.438$ & $0.420$ & $0.502$ & $0.495$ & $0.469$\\
\hline
\multicolumn{5}{l}{ms deviation $\nicefrac{1}{N} \sum_i (\mu_\mathrm{theo} - \mu_\mathrm{exp})^2/ \mu_\mathrm{N}^2$} & $0.062$ & $0.119$ & $0.016$ & $0.031$ & $0.022$ & $0.042$
\end{tabular}}
\flushleft
{\footnotesize $^\text{a}$ recalculated with the $A$ constants from \cite{Sommer.2022} and $\mu\left(^{61}\mathrm{Ni}\right)=-0.7473(4)\,\mu_{\mathrm{N}}$ from \cite{Antusek.2020}}
\end{table*}

\begingroup
\begin{table*}[t]
\centering
\caption{\label{tab:q} Spectroscopic electric quadrupole moments $Q$ of the odd-mass nickel isotopes $^{59,61,65}$Ni determined using Eq.\,\eqref{eq:q} with $^{61}$Ni as the reference isotope. The experimental results are compared to shell-model calculations and \textit{ab initio} VS-IMSRG results, for details see Tab.\,\ref{tab:m}. All values are given in e\,fm$^2$.\\}
\resizebox{\textwidth}{!}{%
\begin{tabular}{r r r r r r r r r r r}
$A$ & $I^\pi$ & This work & Literature & References & \multicolumn{3}{c}{Shell Model}  & \multicolumn{3}{c}{VS-IMSRG} \\
&  &  &  & & \multicolumn{1}{c}{JUN45} & \multicolumn{1}{c}{jj44b} & \multicolumn{1}{c}{GXPF1A} & \multicolumn{1}{c}{1.8/2.0 (EM)} & \multicolumn{1}{c}{$\Delta$N$^2$LO$_\mathrm{GO}$} & \multicolumn{1}{c}{N$^3$LO$_\mathrm{lnl}$} \\
\hline
55 & $\nicefrac{7}{2}^-$ & \multicolumn{1}{c}{--} & \multicolumn{1}{c}{--} & & \multicolumn{1}{c}{--} & \multicolumn{1}{c}{--} & $17.95$ & $14.81$ & $11.24$ & $16.36$ \\
57 & $\nicefrac{3}{2}^-$ & \multicolumn{1}{c}{--} & \multicolumn{1}{c}{--} & & $-7.71$ & $-7.71$ & $-10.85$ & $-8.63$ & $-7.13$ & $-9.79$ \\
59 & $\nicefrac{3}{2}^-$ & $8.8(0.9)$ & \multicolumn{1}{c}{--} & & $8.51$ & $0.31$ & $6.41$ & $-0.80$ & $0.98$ & $0.54$ \\
61 & $\nicefrac{3}{2}^-$ & $16.3(1.5)$ & $16.2(1.5)$ & \cite{Childs.1968, Stone.2014} & $10.16$ & $4.61$ & $17.27$ & $6.23$ & $3.68$ & $8.32$ \\
65 & $\nicefrac{5}{2}^-$ & $9.7(0.9)$ & \multicolumn{1}{c}{--} & & $6.87$ & $10.10$ & $13.33$ & $6.80$ & $7.30$ & $6.70$
\end{tabular}}
\end{table*}

\begingroup
\begin{table}[t]
\centering
\caption{\label{tab:bb} One- and two-body-current contributions to the magnetic dipole moments calculated with the \textit{ab initio} VS-IMSRG, for the results listed in Tab.\,\ref{tab:m} and shown in Fig.\,\ref{fig:moments}b. All values are given in units of the nuclear magneton $\mu_\mathrm{N}$.\\}
\resizebox{\columnwidth}{!}{%
\begin{tabular}{r r r r r r r r}
$A$ & $I^\pi$ & \multicolumn{2}{c}{1.8/2.0 (EM)} & \multicolumn{2}{c}{$\Delta$N$^2$LO$_\mathrm{GO}$} & \multicolumn{2}{c}{N$^3$LO$_\mathrm{lnl}$} \\
 &  & \multicolumn{1}{c}{1-body} & \multicolumn{1}{c}{2-body} & \multicolumn{1}{c}{1-body} & \multicolumn{1}{c}{2-body} & \multicolumn{1}{c}{1-body} & \multicolumn{1}{c}{2-body} \\
\hline
55 & $\nicefrac{7}{2}^-$ & $-0.407$ & $-0.353$ & $-0.344$ & $-0.422$ & $-0.263$ & $-0.391$ \\
57 & $\nicefrac{3}{2}^-$ & $-0.726$ & $-0.117$ & $-0.660$ & $-0.153$ & $-0.466$ & $-0.124$ \\
59 & $\nicefrac{3}{2}^-$ & $-0.808$ & $-0.108$ & $-0.655$ & $-0.138$ & $-0.563$ & $-0.112$ \\
61 & $\nicefrac{3}{2}^-$ & $-0.866$ & $-0.089$ & $-0.791$ & $-0.114$ & $-0.630$ & $-0.093$ \\
63 & $\nicefrac{1}{2}^-$ & $0.649$ & $-0.058$ & $0.577$ & $-0.054$ & $0.691$ & $-0.034$ \\
65 & $\nicefrac{5}{2}^-$ & $1.156$ & $-0.330$ & $1.063$ & $-0.314$ & $1.134$ & $-0.339$ \\
67 & $\nicefrac{1}{2}^-$ & $0.609$ & $-0.107$ & $0.584$ & $-0.089$ & $0.595$ & $-0.126$  
\end{tabular}
}
\flushleft
\end{table}

All values listed in these compilations were still based on the measurements in 1964 \cite{Drain.1964}, even though new measurements of $^{61}\mathrm{Ni(CO)_4}$ in $\mathrm{C_6D_6}$ relative to tetramethylsilane (TMS) were reported in \cite{Harris.2001}. Recently, Antusek \etal\ \cite{Antusek.2020} have performed elaborate \textit{ab initio} calculations of absolute shielding constants of $d$-group elements including the diamagnetic shielding, the chemical shift and solvent effects. The magnitude of the new magnetic moment of $\mu(^{61}\mathrm{Ni})=-0.7473(4)\,\mu_\mathrm{N}$ is again $0.0023\,\mu_\mathrm{N}$ smaller and its estimated uncertainty about an order of magnitude larger than listed in \cite{Stone.2019}. Similar deviations were found for many of the transition group elements, with the largest discrepancies towards the middle of each period~\cite{Antusek.2020}. Here, we use the new value from \cite{Antusek.2020} as the reference for the determination of the magnetic moments of all odd-mass Ni isotopes that were measured. The new reference slightly affects the result for $^{55}$Ni reported recently in \cite{Sommer.2022} but the required correction is negligible compared to its uncertainty, as shown in Tab.\,\ref{tab:m}. In contrast, the magnetic moment of $^{57}$Ni taken from \cite{Ohtsubo.1996} requires no further correction since it has been measured in a nickel crystal and has been extrapolated to a vanishing external field.

Electric quadrupole moments of nuclei are usually much less precise than magnetic dipole moments. For $^{61}$Ni we take the value of $Q=162(15)$\,mb as observed in atomic beam magnetic resonance measurements \cite{Childs.1968} and listed in the latest tabulation of electric quadrupole moments \cite{Pyykko.2018}. This value includes a Sternheimer correction of about 25\% accounting for the redistribution of the inner-shell electrons by the nuclear charge distribution \cite{Sternheimer.1972}. Within the given uncertainty, Eq.\,(\ref{eq:q}) can be expected to be a good approximation to calculate the other quadrupole moments.

The magnetic moments and quadrupole moments obtained are listed in the left part of Tabs.\,\ref{tab:m} and \ref{tab:q}, respectively, and compared to existing values from literature \cite{Stone.2014, Childs.1968, Krane.1976, Rikovska.2000, Dyachkov.2017, Antusek.2020}.  
The dipole moment of $^{59}$Ni and the quadrupole moments of $^{59,65}$Ni have been determined for the first time, while the dipole moments of $^{63,65,67}$Ni agree with literature values but are improved in accuracy by one to two orders of magnitude. Since $A_\mathrm{ref}$ and $B_\mathrm{ref}$ are taken from \cite{Childs.1968}, we can also calculate an experimental value for $^{61}$Ni based on our measurement, which agrees with the reference value since the $A$ and $B$ factors are in excellent agreement. 

\subsection{Nuclear structure calculations}

We have performed \textit{ab initio} VS-IMSRG and phenomenological shell-model calculations and compare them with the experimentally determined nuclear moments. The shell-model calculations were carried out both with \texttt{NuShellX@MSU} \cite{Brown.2014} and \texttt{KSHELL} \cite{KSHELL} and were in good agreement with each other. We explore three different phenomenological interactions, in the following labeled as JUN45 \cite{Honma.2009}, jj44b \cite{Mukhopadhyay.2017}, and GXPF1A \cite{Honma.2005}, which were previously used to calculate the nuclear moments in the neighboring Cu isotopic chain \cite{Vingerhoets.2010, Vingerhoets.2011}. There, the effective $g_\mathrm{eff}$ factors and charges have been optimized to fit the experimental results. In order to evaluate the capability of the shell-model calculations to predict nuclear moments, none of our experimental Ni data was used to set them up, but instead, the same $g_\mathrm{eff}$ factors and charges as well as mass-scaling formulas for the harmonic-oscillator energies used for Cu were applied here. For JUN45 and jj44b, effective $g$-factors and charges of $g^\mathrm{p/n}_{s} = 0.7g^\mathrm{p/n}_{s,\mathrm{free}}$, $g^\mathrm{p}_{l} = 1$, $g^\mathrm{n}_{l} = 0$ and $e^\mathrm{p} = 1.5$, $e^\mathrm{n} = 1.1$ were used with $\hbar\omega = 41 A^{-1/3}$. For GXPF1A, they were set to $g^\mathrm{p/n}_{s} = 0.9g^\mathrm{p/n}_{s,\mathrm{free}}$, $g^\mathrm{p}_{l} = 1.1$, $g^\mathrm{n}_{l} = -0.1$ and $e^\mathrm{p} = 1.5$, $e^\mathrm{n} = 0.5$ with $\hbar\omega = 45 A^{-1/3} - 25 A^{-2/3}$. As a check of consistency, the results presented in \cite{Vingerhoets.2010, Vingerhoets.2011} were reproduced within this work. The chosen interactions are defined in different valence spaces. The JUN45 and jj44b interactions use a closed $^{56}$Ni core such that only the remaining neutrons are free to occupy the $2p_{3/2}$, $1f_{5/2}$, $2p_{1/2}$ and $1g_{9/2}$ single-particle orbitals while the GXPF1A interaction uses a closed $^{40}$Ca core and both the remaining protons and neutrons can occupy the $1f_{7/2}$, $2p_{3/2}$, $1f_{5/2}$ and $2p_{1/2}$ shells. Calculations including both the $1f_{7/2}$ and $1g_{9/2}$ shell without any further restrictions exceed the current computational limits. It should be noted that the phenomenological interactions used are fitted to nuclear energies extracted from nuclear spectroscopy data. More details can be found in the respective references.

\begin{figure*}[t]
\centering
\includegraphics[width=\linewidth]{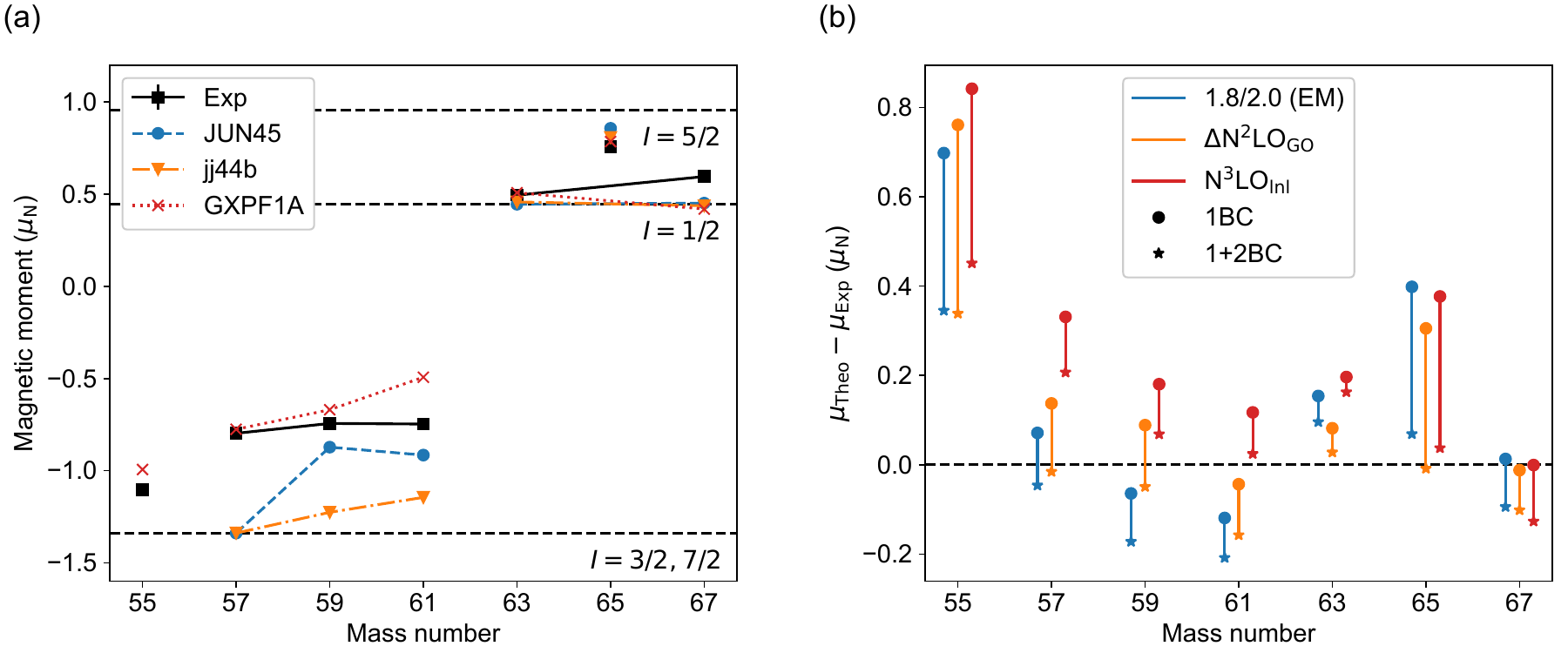}
\caption{\label{fig:moments} (a) Comparison of the nuclear magnetic moments from shell-model calculations with the experimental results. The dashed lines depict the effective Schmidt values for the different nuclear spins $I$.  (b) Difference between the nuclear magnetic moments obtained in \textit{ab initio} VS-IMSRG calculations, with and without two-body currents, and experimental values. The experimental result of $^{55}$Ni is taken from \cite{Sommer.2022}, that of $^{57}$Ni from \cite{Ohtsubo.1996}.}
\end{figure*}

The VS-IMSRG~\cite{Hergert.2015, Stroberg2019} calculations were performed with the nucleon-nucleon (NN) and three-nucleon (3N) interactions, and electromagnetic currents based on chiral effective field theory (EFT).
In this work, we used three established NN+3N interactions 1.8/2.0 (EM)~\cite{Hebeler2011}, N$^{3}$LO$_{\rm lnl}$~\cite{Soma2020}, and $\Delta$N$^{2}$LO$_{\rm GO}$~\cite{Jiang2020}.
The 1.8/2.0 (EM) interaction is composed of the NN interaction up to next-to-next-to-next-to leading order (N$^{3}$LO)~\cite{Entem.2003} softened by the free-space similarity renormalization group (SRG) to momentum scale 1.8\,fm$^{-1}$ and 3N interaction at N$^{2}$LO regulated a 2.0\,fm$^{-1}$ cutoff regulator function. The NN part is fitted to NN scattering data, and the two 3N couplings to the $^{3}$H binding energy and the $^{4}$He radius.
The N$^{3}$LO$_{\rm lnl}$ interaction includes the NN interaction~\cite{Entem.2003} and local-non-locally regulated 3N interaction~\cite{Soma2020} fitted to $A=2, 3$, and $4$ systems. 
In this work, the NN+3N interaction is consistently softened by the SRG technique to the momentum scale 2.0\,fm$^{-1}$ to accelerate the convergence of many-body calculations. 
Finally, we employed the recently developed $\Delta$N$^{2}$LO$_{\rm GO}$~\cite{Jiang2020} interaction based on $\Delta$-full chiral EFT with cutoff $394$ MeV, which is fit to few-body data, light nuclei, and nuclear matter properties.
For the magnetic dipole operator, in addition to the point-like nucleon contributions, we include the leading 2BCs in the chiral EFT expansion~\cite{Pastore2009,Krebs2019}, whose matrix elements are computed as given in Ref.~\cite{Seutin2023}.
The calculations begin with the operators expressed in the 13 major-shell harmonic-oscillator (HO) space combined with the recently introduced 3N storage scheme~\cite{Miyagi2022}, enabling us to use a sufficiently large truncation for the 3N matrix elements.
After the Hartree-Fock calculation using the ensemble normal ordering technique~\cite{Stroberg2017}, the $pf$-shell valence space is decoupled from the rest of the Hilbert space truncating the VS-IMSRG transformation at the normal-ordered two-body level using the Magnus expansion~\cite{Morris2015}.
The one- and two-body magnetic dipole operators are transformed in the same manner~\cite{Parzuchowski2017}.
Note that we do not use any effective parameters in the VS-IMSRG calculations.

\section{Discussion}
\label{sec:discussion}

The experimental magnetic moments (black squares) are plotted in Fig.\,\ref{fig:moments}(a) together with the shell-model predictions and the effective Schmidt values (dashed lines).  The isotopes $^{63,65,67}$Ni having nuclear spins of $\nicefrac{1}{2}$ and $\nicefrac{5}{2}$ are fairly close to the respective effective Schmidt values. At $N=40$ the negative parity states are filled, while the higher-lying $\nicefrac{9}{2}^+$ states have opposite parity and are not accessible by M1 transitions. Thus, one expects this behavior close to $N=40$. The shell-model calculations are quite similar for all model spaces and relatively close to the experimental data for $^{63,65}$Ni. For the $\nicefrac{1}{2}^+$ states they all predict the nucleus to have essentially a magnetic moment close to the effective Schmidt value. While for the neutron-rich nuclei, the JUN45 and jj44b interactions result in wave functions with some population in the $g_{\nicefrac{9}{2}}$ states, the GXPF1A interaction sorts this population mostly into the $f_{\nicefrac{5}{2}}$ state. 
Since the resulting magnetic moments for $^{63,65,67}$Ni are quite similar, it can be concluded that for these isotopes, neither the inclusion of the $f_{\nicefrac{7}{2}}$ nor of the $g_{\nicefrac{9}{2}}$ states brings much of a benefit. However, it should be noted that the quadrupole moment of $^{65}$Ni varies by about a factor of two between the different shell-modell calculations and is for jj44b closest to the experimental value, while for JUN45 it is comparable to the smaller values obtained in all \textit{ab initio} calculations.
Contrary, the magnetic moments of the lighter isotopes $^{57,59,61}$Ni ($I=\nicefrac{3}{2}$) deviate strongly from the single-particle expectation and the shell-model calculations become much more sensitive to the employed interactions, even though approaching the doubly magic $^{56}$Ni. We note that neither the magnetic moment of $^{57}$Ni \cite{Ohtsubo.1996} nor that of $^{55}$Ni are close to the single-particle value \cite{Sommer.2022}, indicating that the doubly magic $^{56}$Ni is a rather soft core and particle-hole excitations across $N=28$ might considerably contribute to its ground state. 
From this point of view it is not surprising that calculations using the GXPF1A interaction are closest to the experimental value of $^{57,59}$Ni. Note that both JUN45 and jj44b are not available for $^{55}$Ni and yield the effective Schmidt value for $^{57}$Ni as there is only one valence nucleon. Against the intuition, the interaction JUN45 fitted to binding and excitation energies of nuclei with proton and neutron numbers outside of the range of investigated nickel isotopes yields a considerably better agreement with experiment than jj44b, where information from nickel isotopes has additionally been used in the optimization.

Figure\,\ref{fig:moments}(b) shows the comparison to the VS-IMSRG results. Including only one-body currents (1BC), the \textit{ab initio} calculations are already in agreement with the experimental data to a level that is on a par with the best shell-model results for $N > 28$. Expressed as mean-square (ms) deviation $\nicefrac{1}{N} \sum_i (\mu_i^\mathrm{theo} - \mu_{i}^\mathrm{exp})^2$ from the experimental results, above the shell closure, $\Delta$N$^{2}$LO$_{\rm GO}$ ($0.021\,\mu_\mathrm{N}^2$) including only 1BC is close to the best shell model results provided by GXPF1A ($0.017\,\mu_\mathrm{N}^2$). Below the shell closure, the magnetic moment of $^{55}$Ni is substantially underestimated by the 1BC, leading to an overall large ms deviation. As listed in Tab.\,\ref{tab:bb}, the 2BC contributions are always negative. For the N$^3$LO$_\mathrm{lnl}$ interaction, they bring the magnetic moment in all cases with the exception of $^{67}$Ni closer to the experimental data, see Fig.\,\ref{fig:moments}(b).
The inclusion of 2BCs reduces the ms deviation from the experimental data for all interactions significantly. For the $\Delta$N$^{2}$LO$_{\rm GO}$ interaction it is improved by a factor of 5 from 0.101\,$\mu_\mathrm{N}^2$ to 0.022\,$\mu_\mathrm{N}^2$, which brings it on par with the ms deviation of the GXPF1A result if also $^{55}$Ni is included and is now almost a factor of 3 smaller if only the isotopes above $N=28$ are considered. The improvement factor for EM\,1.8/2.0 and N$^3$LO$_\mathrm{lnl}$ with respect to the 1BC calculation is approximately 3.

The largest 2BC contributions appear for $^{55,65}$Ni ($I=\nicefrac{7}{2}^-, \nicefrac{5}{2}^-$) and shift the moments substantially closer to the experimental results for all used interactions. This is consistent with the finding of~\cite{Miyagi2023}, \textit{i.e.}, the 2BC contribution tends to be larger if the last unpaired nucleon is more spatially extended.
Since we observed that the largest configuration for the last unpaired neutron is the $l=3$ orbital for $^{55,65}$Ni and $l=1$ for $^{57,59,61,63,67}$Ni, the last unpaired neutron distributions in $^{55,65}$Ni are expected to be broader than those in the other isotopes. These corrections clearly dominate the overall reduction of the ms deviation. Similar to the shell-model calculations, the interaction dependence (uncertainty) is significantly larger for $^{57,59,61}$Ni as these isotopes are closer to the soft shell closure at $N=28$.

The newly measured spectroscopic quadrupole moments of $^{59,65}$Ni are both approximately half as large as that of the reference nucleus $^{61}$Ni. This corresponds to the trend of the $B$(E2) values in the respective even-even isotopes, which rise from $^{56}$Ni to a maximum around $^{60,62}$Ni before they decrease again~\cite{Pritychenko.2012} towards the sub-shell closure at $^{68}$Ni. 
Of all shell-model calculations, this is reproduced best with the GXPF1A interaction, including excitations from the $f_{\nicefrac{7}{2}}$ proton and neutron levels. 
The VS-IMSRG calculations with the N$^3$LO$_\mathrm{lnl}$ interaction reproduce the general trend best, even though the magnitudes are considerably too small.
All calculations indicate a change from prolate to oblate deformation when going from a single-hole state in $^{55}$Ni to a single-particle state in $^{57}$Ni as expected in a single-particle model \cite{Goeppert-Mayer.1955}. 
Note that vector 2BC also contribute to electric quarupole moments and E2 observables. However, while the contribution of 2BC to magnetic moments were benchmarked in recent works \cite{Seutin2023, Miyagi2023}, systematic tests of E2 2BC contributions for use in \textit{ab initio} VS-IMSRG calculations have not been carried out yet.

\section{Summary}

We measured the nuclear moments of the even-odd isotopes $^{59-67}$Ni and compared them to phenomenological shell-model and \textit{ab initio} valence-space in-medium similarity renormalization group (VS-IMSRG) calculations. For the latter, two-body current (2BC) contributions were included and improved the predictive power of the calculation in terms of the mean-square deviation over these isotopes. The best overall agreement with the experimental data was found for the VS-IMSRG calculations using the $\Delta$N$^2$LO$_\mathrm{GO}$ interaction, whose mean-square deviation across all isotopes is on par with the best shell-model calculations, which were obtained with the GXPF1A interaction.

Our results constitute an important benchmark for VS-IMSRG calculations including 2BCs, which can be used to predict the structure of nuclei that are out of reach of laboratory experiments. This is especially important where the r-process takes place \cite{Schatz.2022}.
The outstanding role of nickel as the heaviest element produced in the burning of stars, its interesting nuclear structure and the remaining deviation with the experiment suggest reevaluating the nickel moments with higher-order VS-IMSRG calculations in the future.
From the experimental side, an extension of the measurements to $A \geq 69$ would be useful to test calculations towards neutron-rich extremes.

\section*{Acknowledgements}

We acknowledge the support of the ISOLDE Collaboration and technical teams. This work was supported by the European Research Council (ERC) under the European Union's Horizon 2020 research and innovation programme (Grant Agreements No.~654002 and No.~101020842), the Deutsche Forschungsgemeinschaft (DFG, German Research Foundation) -- Project-ID 279384907 -- SFB 1245, the BMBF under Contracts No.~05P18RDCIA, 05P19RDFN1, and 05P21RDCI1, the FWO (Belgium), GOA 15/010 from KU Leuven, and by consolidated grants from STFC (UK) - ST/L005670/1, ST/L005794/1, ST/P004423/1, and ST/P004598/1.
For the VS-IMSRG calculations, the \texttt{NuHamil}~\cite{NuHamil}, \texttt{IMSRG++}~\cite{imsrg++}, and \texttt{KSHELL}~\cite{KSHELL} codes were used to generate chiral EFT matrix elements, to perform the valence-space decoupling, and to solve the valence-space problems, respectively.

\bibliographystyle{elsarticle-num-title-linked}
\bibliography{main}

\begin{thebibliography}{10}
\expandafter\ifx\csname url\endcsname\relax
  \def\url#1{\texttt{#1}}\fi
\expandafter\ifx\csname urlprefix\endcsname\relax\def\urlprefix{URL }\fi
\expandafter\ifx\csname href\endcsname\relax
  \def\href#1#2{#2} \def\path#1{#1}\fi

\bibitem{Schatz.2022}
{H.\ Schatz \textit{et al.}},
  \href{https://doi.org/10.1088/1361-6471/ac8890}{{Horizons: Nuclear
  astrophysics in the 2020s and beyond}}, J. Phys. G 49 (2022) 110502.

\bibitem{Truran.1967}
J.~W. Truran, W.~D. Arnett, A.~G.~W. Cameron,
  \href{https://doi.org/10.1139/p67-184}{{Nucleosynthesis in supernova shock
  waves}}, Can. J. Phys. 45 (1967) 2315--2332.

\bibitem{Timmes.2003}
F.~X. Timmes, E.~F. Brown, J.~W. Truran,
  \href{https://dx.doi.org/10.1086/376721}{{On Variations in the Peak
  Luminosity of Type Ia Supernovae}}, Astrophys. J. 590 (2003) L83.

\bibitem{Chieffi.2017}
A.~Chieffi, M.~Limongi,
  \href{https://dx.doi.org/10.3847/1538-4357/836/1/79}{{The Synthesis of
  $^{44}$Ti and $^{56}$Ni in Massive Stars}}, Astrophys. J. 836 (2017) 79.

\bibitem{Thielemann.2018}
F.-K. Thielemann, J.~Isern, A.~Perego, P.~von Ballmoos,
  \href{https://doi.org/10.1007/s11214-018-0494-5}{{Nucleosynthesis in
  Supernovae}}, Space Sci. Rev. 214 (2018) 62.

\bibitem{Sommer.2022}
F.~Sommer, K.~K{\"o}nig, D.~M. Rossi, N.~Everett, D.~Garand, R.~P. de~Groote,
  J.~D. Holt, P.~Imgram, A.~Incorvati, C.~Kalman, A.~Klose, J.~Lantis, Y.~Liu,
  A.~J. Miller, K.~Minamisono, T.~Miyagi, W.~Nazarewicz,
  W.~N{\"o}rtersh{\"a}user, S.~V. Pineda, R.~Powel, P.-G. Reinhard, L.~Renth,
  E.~Romero-Romero, R.~Roth, A.~Schwenk, C.~Sumithrarachchi,
  A.~Teigelh{\"o}fer,
  \href{https://doi.org/10.1103/PhysRevLett.129.132501}{{Charge Radii of
  $^{55,56}$Ni Reveal a Surprisingly Similar Behavior at $N$=28 in Ca and Ni
  Isotopes}}, Phys. Rev. Lett. 129 (2022) 132501.

\bibitem{Malbrunot-Ettenauer.2022}
S.~Malbrunot-Ettenauer, S.~Kaufmann, S.~Bacca, C.~Barbieri, J.~Billowes, M.~L.
  Bissell, K.~Blaum, B.~Cheal, T.~Duguet, R.~F.~G. Ruiz, W.~Gins, C.~Gorges,
  G.~Hagen, H.~Heylen, J.~D. Holt, G.~R. Jansen, A.~Kanellakopoulos,
  M.~Kortelainen, T.~Miyagi, P.~Navr\'atil, W.~Nazarewicz, R.~Neugart,
  G.~Neyens, W.~N\"ortersh\"auser, S.~J. Novario, T.~Papenbrock, T.~Ratajczyk,
  P.-G. Reinhard, L.~V. Rodr\'{\i}guez, R.~S\'anchez, S.~Sailer, A.~Schwenk,
  J.~Simonis, V.~Som\`a, S.~R. Stroberg, L.~Wehner, C.~Wraith, L.~Xie, Z.~Y.
  Xu, X.~F. Yang, D.~T. Yordanov,
  \href{https://link.aps.org/doi/10.1103/PhysRevLett.128.022502}{{Nuclear
  Charge Radii of the Nickel Isotopes $^{58-68,70}\mathrm{Ni}$}}, Phys. Rev.
  Lett. 128 (2022) 022502.

\bibitem{Pineda.2021}
S.~V. Pineda, K.~K{\"o}nig, D.~M. Rossi, B.~A. Brown, A.~Incorvati, J.~Lantis,
  K.~Minamisono, W.~N{\"o}rtersh{\"a}user, J.~Piekarewicz, R.~Powel, F.~Sommer,
  \href{https://doi.org/10.1103/PhysRevLett.127.182503}{{Charge Radius of
  Neutron-Deficient $^{54}$Ni and Symmetry Energy Constraints Using the
  Difference in Mirror Pair Charge Radii}}, Phys. Rev. Lett. 127 (2021) 182503.

\bibitem{Kaufmann.2020}
S.~Kaufmann, J.~Simonis, S.~Bacca, J.~Billowes, M.~L. Bissell, K.~Blaum,
  B.~Cheal, R.~F.~G. Ruiz, W.~Gins, C.~Gorges, G.~Hagen, H.~Heylen,
  A.~Kanellakopoulos, S.~Malbrunot-Ettenauer, M.~Miorelli, R.~Neugart,
  G.~Neyens, W.~N\"ortersh\"auser, R.~S\'anchez, S.~Sailer, A.~Schwenk,
  T.~Ratajczyk, L.~V. Rodr\'{\i}guez, L.~Wehner, C.~Wraith, L.~Xie, Z.~Y. Xu,
  X.~F. Yang, D.~T. Yordanov,
  \href{https://link.aps.org/doi/10.1103/PhysRevLett.124.132502}{{Charge Radius
  of the Short-Lived $^{68}$Ni and Correlation with the Dipole
  Polarizability}}, Phys. Rev. Lett. 124 (2020) 132502.

\bibitem{Neugart.1981}
R.~Neugart, \href{https://doi.org/10.1016/0029-554X(81)90902-2}{{Laser
  spectroscopy on mass-separated radioactive beams}}, Nucl. Instrum. Methods
  Phys. Res. 186 (1981) 165--175.

\bibitem{Otten.1989}
E.~W. Otten, \href{https://doi.org/10.1007/978-1-4613-0713-6_7}{{Nuclear Radii
  and Moments of Unstable Isotopes}}, in: D.~A. Bromley (Ed.), Treatise on
  Heavy Ion Science, {Springer Boston}, 1989, pp. 517--638.

\bibitem{Neyens.2003}
G.~Neyens, \href{https://dx.doi.org/10.1088/0034-4885/66/4/205}{Nuclear
  magnetic and quadrupole moments for nuclear structure research on exotic
  nuclei}, Rep. Prog. Phys. 66 (2003) 633.

\bibitem{Campbell.2016}
P.~Campbell, I.~D. Moore, M.~R. Pearson,
  \href{https://doi.org/10.1016/j.ppnp.2015.09.003}{{Laser spectroscopy for
  nuclear structure physics}}, Prog. Part. Nucl. Phys. 86 (2016) 127--180.

\bibitem{Schmidt.1937}
T.~Schmidt, \href{https://doi.org/10.1007/BF01338744}{{{\"U}ber die
  magnetischen Momente der Atomkerne}}, Z. Phys. 106 (1937) 358--361.

\bibitem{Andl.1982}
A.~Andl, K.~Bekk, S.~G\"oring, A.~Hanser, G.~Nowicki, H.~Rebel, G.~Schatz,
  R.~C. Thompson, \href{https://doi.org/10.1103/PhysRevC.26.2194}{Isotope
  shifts and hyperfine structure of the $4s^2\, ^1{S}_0-4s4p\, ^1{P}_1$
  transition in calcium isotopes}, Phys. Rev. C 26 (1982) 2194--2202.

\bibitem{Ohtsubo.1996}
T.~Ohtsubo, D.~J. Cho, Y.~Yanagihashi, S.~Ohya, S.~Muto,
  \href{https://doi.org/10.1103/PhysRevC.54.554}{{Measurement of the nuclear
  magnetic moments of $^{57}$Ni and $^{59}$Fe}}, Phys. Rev. C 54 (1996)
  554--558.

\bibitem{Yordanov.2020}
D.~T. Yordanov, L.~V. Rodr{\'i}guez, D.~L. Balabanski, J.~Biero{\'{n}}, M.~L.
  Bissell, K.~Blaum, B.~Cheal, J.~Ekman, G.~Gaigalas, R.~F. Garcia~Ruiz,
  G.~Georgiev, W.~Gins, M.~R. Godefroid, C.~Gorges, Z.~Harman, H.~Heylen,
  P.~J{\"o}nsson, A.~Kanellakopoulos, S.~Kaufmann, C.~H. Keitel, V.~Lagaki,
  S.~Lechner, B.~Maa{\ss}, S.~Malbrunot-Ettenauer, W.~Nazarewicz, R.~Neugart,
  G.~Neyens, W.~N{\"o}rtersh{\"a}user, N.~S. Oreshkina, A.~Papoulia,
  P.~Pyykk{\"o}, P.-G. Reinhard, S.~Sailer, R.~S{\'a}nchez, S.~Schiffmann,
  S.~Schmidt, L.~Wehner, C.~Wraith, L.~Xie, Z.~Xu, X.~Yang,
  \href{https://doi.org/10.1038/s42005-020-0348-9}{Structural trends in atomic
  nuclei from laser spectroscopy of tin}, Commun. Phys. 3 (2020) 107.

\bibitem{Castel.1990}
B.~Castel, I.~S. Towner,
  \href{https://doi.org/10.1093/oso/9780198517283.001.0001}{{Modern Theories of
  Nuclear Moments}}, Oxford University Press, 1990.

\bibitem{Brown.2001}
B.~Brown,
  \href{https://www.sciencedirect.com/science/article/pii/S0146641001001594}{The
  nuclear shell model towards the drip lines}, Prog. Part. Nucl. Phys. 47
  (2001) 517--599.

\bibitem{Vingerhoets.2010}
P.~Vingerhoets, K.~T. Flanagan, M.~Avgoulea, J.~Billowes, M.~L. Bissell,
  K.~Blaum, B.~A. Brown, B.~Cheal, M.~de~Rydt, D.~H. Forest, C.~Geppert,
  M.~Honma, M.~Kowalska, J.~Kr{\"a}mer, A.~Krieger, E.~Mane, R.~Neugart,
  G.~Neyens, W.~N{\"o}rtersh{\"a}user, T.~OtsukA, M.~Schug, H.~H. Stroke,
  G.~Tungate, D.~T. Yordanov,
  \href{https://doi.org/10.1103/PhysRevC.82.064311}{{Nuclear spins, magnetic
  moments, and quadrupole moments of Cu isotopes from $N$=28 to $N$=46: Probes
  for core polarization effects}}, Phys. Rev. C 82 (2010) 064311.

\bibitem{Kanellakopoulos.2020}
A.~Kanellakopoulos, X.~F. Yang, M.~L. Bissell, M.~L. Reitsma, S.~W. Bai,
  J.~Billowes, K.~Blaum, A.~Borschevsky, B.~Cheal, C.~S. Devlin, R.~F. {Garcia
  Ruiz}, H.~Heylen, S.~Kaufmann, K.~K{\"o}nig, {\'A}.~Koszor{\'u}s, S.~Lechner,
  S.~Malbrunot-Ettenauer, R.~Neugart, G.~Neyens, W.~N{\"o}rtersh{\"a}user,
  T.~Ratajczyk, L.~V. Rodr\'{\i}guez, S.~Sels, S.~J. Wang, L.~Xie, Z.~Y. Xu,
  D.~T. Yordanov, \href{https://doi.org/10.1103/PhysRevC.102.054331}{{Nuclear
  moments of germanium isotopes near $N=40$}}, Phys. Rev. C 102 (2020) 054331.

\bibitem{Lechner.2023}
S.~Lechner, T.~Miyagi, Z.~Y. Xu, M.~L. Bissell, K.~Blaum, B.~Cheal, C.~S.
  Devlin, R.~F. {Garcia Ruiz}, J.~S.~M. Ginges, H.~Heylen, J.~D. Holt,
  P.~Imgram, A.~Kanellakopoulos, A.~Koszorús, S.~Malbrunot-Ettenauer,
  R.~Neugart, G.~Neyens, W.~N\"ortersh\"auser, P.~Plattner, L.~V.
  Rodr\'{\i}guez, G.~Sanamyan, S.~R. Stroberg, Y.~Utsuno, X.~F. Yang, D.~T.
  Yordanov,
  \href{https://doi.org/10.1016/j.physletb.2023.138278}{{Electromagnetic
  moments of the antimony isotopes $^{112-133}$Sb}}, Phys. Lett. B 847 (2023)
  138278.

\bibitem{Gysbers2019}
P.~Gysbers, G.~Hagen, J.~D. Holt, G.~R. Jansen, T.~D. Morris,
  P.~Navr{\'{a}}til, T.~Papenbrock, S.~Quaglioni, A.~Schwenk, S.~R. Stroberg,
  K.~A. Wendt, \href{https://doi.org/10.1038/s41567-019-0450-7}{{Discrepancy
  between experimental and theoretical $\beta$-decay rates resolved from first
  principles}}, Nat. Phys. 15 (2019) 428--431.

\bibitem{Pastore2013}
S.~Pastore, S.~C. Pieper, R.~Schiavilla, R.~B. Wiringa,
  \href{https://link.aps.org/doi/10.1103/PhysRevC.87.035503}{{Quantum Monte
  Carlo calculations of electromagnetic moments and transitions in $A \leq 9$
  nuclei with meson-exchange currents derived from chiral effective field
  theory}}, Phys. Rev. C 87 (2013) 035503.

\bibitem{Friman-Gayer2021}
U.~Friman-Gayer, C.~Romig, T.~H{\"{u}}ther, K.~Albe, S.~Bacca, T.~Beck,
  M.~Berger, J.~Birkhan, K.~Hebeler, O.~J. Hernandez, J.~Isaak, S.~K{\"{o}}nig,
  N.~Pietralla, P.~C. Ries, J.~Rohrer, R.~Roth, D.~Savran, M.~Scheck,
  A.~Schwenk, R.~Seutin, V.~Werner,
  \href{https://doi.org/10.1103/PhysRevLett.126.102501}{{Role of Chiral
  Two-Body Currents in $^{6}\mathrm{Li}$ Magnetic Properties in Light of a New
  Precision Measurement with the Relative Self-Absorption Technique}}, Phys.
  Rev. Lett. 126 (2021) 102501.

\bibitem{Acharya2023}
B.~Acharya, B.~S. Hu, S.~Bacca, G.~Hagen, P.~Navr\'atil, T.~Papenbrock,
  \href{https://doi.org/10.48550/arXiv.2311.11438}{{The magnetic dipole
  transition in $^{48}$Ca}}, arXiv:2311.11438 (2023).

\bibitem{Miyagi2023}
T.~Miyagi, X.~Cao, R.~Seutin, S.~Bacca, R.~F.~G. Ruiz, K.~Hebeler, J.~D. Holt,
  A.~Schwenk, \href{https://doi.org/10.48550/arXiv.2311.14383}{{Impact of
  two-body currents on magnetic dipole moments of nuclei}}, arXiv:2311.14383
  (2023).

\bibitem{Neugart.2017}
R.~Neugart, J.~Billowes, M.~L. Bissell, K.~Blaum, B.~Cheal, K.~T. Flanagan,
  G.~Neyens, W.~Nörtershäuser, D.~T. Yordanov,
  \href{https://doi.org/10.1088/1361-6471/aa6642}{{Collinear laser spectroscopy
  at {ISOLDE}: new methods and highlights}}, J. Phys. G 44 (2017) 064002.

\bibitem{Marsh.2014}
B.~A. Marsh, \href{https://doi.org/10.1063/1.4858015}{{Resonance ionization
  laser ion sources for on-line isotope separators (invited)}}, Rev. Sci.
  Instrum. 85 (2014) 02B923.

\bibitem{Franberg.2008}
H.~Fr{\aa}nberg, P.~Delahaye, J.~Billowes, K.~Blaum, R.~Catherall, F.~Duval,
  O.~Gianfrancesco, T.~Giles, A.~Jokinen, M.~Lindroos, D.~Lunney, E.~Mane,
  I.~Podadera, \href{https://doi.org/10.1016/j.nimb.2008.05.097}{{Off-line
  commissioning of the ISOLDE cooler}}, Nucl. Instrum. Methods Phys. Res. B 266
  (2008) 4502--4504.

\bibitem{Mueller.1983}
A.~C. Mueller, F.~Buchinger, W.~Klempt, E.~W. Otten, R.~Neugart,
  C.~Ekstr{\"o}m, J.~Heinemeier,
  \href{https://doi.org/10.1016/0375-9474(83)90226-9}{{Spins, moments and
  charge radii of barium isotopes in the range $^{122\ensuremath{-}146}$Ba
  determined by collinear fast-beam laser spectroscopy}}, Nucl. Phys. A 403
  (1983) 234--262.

\bibitem{Klose.2012}
A.~Klose, K.~Minamisono, C.~Geppert, N.~Fr{\"o}mmgen, M.~Hammen, J.~Kr{\"a}mer,
  A.~Krieger, C.~D.~P. Levy, P.~F. Mantica, W.~N{\"o}rtersh{\"a}user,
  S.~Vinnikova, \href{https://doi.org/10.1016/j.nima.2012.03.006}{{Tests of
  atomic charge-exchange cells for collinear laser spectroscopy}}, Nucl.
  Instrum. Methods Phys. Res. A 678 (2012) 114--121.

\bibitem{Ryder.2015}
C.~A. Ryder, K.~Minamisono, H.~B. Asberry, B.~Isherwood, P.~F. Mantica,
  A.~Miller, D.~M. Rossi, R.~Strum,
  \href{https://doi.org/10.1016/j.sab.2015.08.004}{{Population distribution
  subsequent to charge exchange of 29.85\,keV Ni$^+$ on sodium vapor}},
  Spectrochim. Acta B 113 (2015) 16--21.

\bibitem{Kreim.2014}
K.~Kreim, M.~L. Bissell, J.~Papuga, K.~Blaum, M.~de~Rydt, R.~F. {Garcia Ruiz},
  S.~Goriely, H.~Heylen, M.~Kowalska, R.~Neugart, G.~Neyens,
  W.~N{\"o}rtersh{\"a}user, M.~M. Rajabali, R.~{S{\'a}nchez Alarc{\'o}n}, H.~H.
  Stroke, D.~T. Yordanov,
  \href{https://doi.org/10.1016/j.physletb.2014.02.012}{{Nuclear charge radii
  of potassium isotopes beyond $N$=28}}, Phys. Lett. B 731 (2014) 97--102.

\bibitem{Verlinde.2020}
M.~Verlinde, K.~Dockx, S.~Geldhof, K.~K{\"o}nig, D.~Studer, T.~E. Cocolios,
  R.~P. de~Groote, R.~Ferrer, Y.~Kudryavtsev, T.~Kieck, I.~Moore,
  W.~N{\"o}rtersh{\"a}user, S.~Raeder, P.~{van den Bergh}, P.~{van Duppen},
  K.~Wendt, \href{https://doi.org/10.1007/s00340-020-07425-4}{{On the
  performance of wavelength meters: Part 1 -- consequences for
  medium-to-high-resolution laser spectroscopy}}, Appl. Phys. B 126 (2020) 85.

\bibitem{Konig.2020}
K.~K{\"o}nig, P.~Imgram, J.~Kr{\"a}mer, B.~Maa{\ss}, K.~Mohr, T.~Ratajczyk,
  F.~Sommer, W.~N{\"o}rtersh{\"a}user,
  \href{https://doi.org/10.1007/s00340-020-07433-4}{{On the performance of
  wavelength meters: Part 2 -- frequency-comb based characterization for more
  accurate absolute wavelength determinations}}, Appl. Phys. B 126 (2020) 86.

\bibitem{Kaufmann.2019}
S.~Kaufmann, \href{http://tuprints.ulb.tu-darmstadt.de/9286/}{{Laser
  spectroscopy of nickel isotopes with a new data acquisition system at
  ISOLDE}}, Ph.D. thesis, {Technische Universit{\"a}t}, Darmstadt (2019).

\bibitem{scipy}
P.~Virtanen, R.~Gommers, T.~E. Oliphant, M.~Haberland, T.~Reddy, D.~Cournapeau,
  E.~Burovski, P.~Peterson, W.~Weckesser, J.~Bright, S.~J. {van der Walt},
  M.~Brett, J.~Wilson, K.~J. Millman, N.~Mayorov, A.~R.~J. Nelson, E.~Jones,
  R.~Kern, E.~Larson, C.~J. Carey, {\.I}.~Polat, Y.~Feng, E.~W. Moore,
  J.~{VanderPlas}, D.~Laxalde, J.~Perktold, R.~Cimrman, I.~Henriksen, E.~A.
  Quintero, C.~R. Harris, A.~M. Archibald, A.~H. Ribeiro, F.~Pedregosa, P.~{van
  Mulbregt}, {SciPy 1.0 Contributors},
  \href{https://doi.org/10.1038/s41592-019-0686-2}{{{SciPy} 1.0: Fundamental
  Algorithms for Scientific Computing in Python}}, Nat. Methods 17 (2020)
  261--272.

\bibitem{Bendali.1986}
N.~Bendali, H.~T. Duong, P.~Juncar, J.~M. Stjalm, J.~L. Vialle,
  \href{https://doi.org/10.1088/0022-3700/19/2/012}{{Na$^+$-Na Charge-Exchange
  Processes Studied by Collinear Laser Spectroscopy}}, J. Phys. B. 19 (1986)
  233--238.

\bibitem{Krieger.2016}
A.~Krieger, W.~N{\"o}rtersh{\"a}user, C.~Geppert, K.~Blaum, M.~L. Bissell,
  N.~Fr{\"o}mmgen, M.~Hammen, K.~Kreim, M.~Kowalska, J.~Kr{\"a}mer, R.~Neugart,
  G.~Neyens, R.~S{\'a}nchez, D.~Tiedemann, D.~T. Yordanov, M.~Zakova,
  \href{https://doi.org/10.1007/s00340-016-6579-5}{{Frequency-comb referenced
  collinear laser spectroscopy of Be$^+$ for nuclear structure investigations
  and many-body QED tests}}, Appl. Phys. B 123 (2016) 15.

\bibitem{NIST}
A.~Kramida, Y.~Ralchenko, J.~Reader, {NIST ASD Team},
  \href{https://doi.org/10.18434/T4W30F}{{NIST Atomic Spectra Database (ver.
  5.10)}}, {National Institute of Standards and Technology, Gaithersburg, MD}
  (2023).

\bibitem{Konig.2021b}
K.~K{\"o}nig, F.~Sommer, J.~Lantis, K.~Minamisono, W.~N{\"o}rtersh{\"a}user,
  S.~Pineda, R.~Powel,
  \href{https://doi.org/10.1103/PhysRevC.103.054305}{{Isotope-shift
  measurements and King-fit analysis in nickel isotopes}}, Phys. Rev. C 103
  (2021) 054305.

\bibitem{Childs.1968}
W.~J. Childs, L.~S. Goodman,
  \href{https://doi.org/10.1103/PhysRev.170.136}{{Hyperfine-Structure Studies
  of $^{61}$Ni, and the Nuclear Ground-State Electric Quadrupole Moment}},
  Phys. Rev. 170 (1968) 136--140.

\bibitem{Stone.2019}
N.~J. Stone,
  \href{http://inis.iaea.org/search/search.aspx?orig_q=RN:51052833}{{Table of
  recommended nuclear magnetic dipole moments}}, Tech. rep., International
  Atomic Energy Agency (IAEA), {Nuclear Physics and Radiation Physics,
  INDC(NDS)--0794} (2019).

\bibitem{Dyachkov.2017}
A.~B. D'yachkov, V.~A. Firsov, A.~A. Gorkunov, A.~V. Labozin, S.~M. Mironov,
  E.~E. Saperstein, S.~V. Tolokonnikov, G.~O. Tsvetkov, V.~Y. Panchenko,
  \href{https://doi.org/10.1140/epja/i2017-12197-5}{{Hyperfine structure of
  electronic levels and the first measurement of the nuclear magnetic moment of
  $^{63}$Ni}}, Eur. J. Phys. A 53 (2017) 13.

\bibitem{Fuller.1969}
G.~H. Fuller, V.~W. Cohen, \href{https://www.osti.gov/biblio/4801983}{{Nuclear
  Spins and Moments}}, Nucl. Data Sect. A 5 (1969) 433--612.

\bibitem{Fuller.1976}
G.~H. Fuller, \href{https://doi.org/10.1063/1.555544}{{Nuclear Spins and
  Moments}}, J. Phys. Chem. Ref. Data 5 (1976) 835--1092.

\bibitem{Dickinson.1950}
W.~C. Dickinson, \href{https://doi.org/10.1103/PhysRev.80.563}{{Hartree
  Computation of the Internal Diamagnetic Field for Atoms}}, Phys. Rev. 80
  (1950) 563--566.

\bibitem{Raghavan.1989}
P.~Raghavan, \href{https://doi.org/10.1016/0092-640X(89)90008-9}{{Table of
  Nuclear Moments}}, At. Data Nucl. Data Tables 42 (1989) 189--291.

\bibitem{Johnson.1983}
W.~R. Johnson, D.~Kolb, K.-N. Huang,
  \href{https://doi.org/10.1016/0092-640X(83)90020-7}{{Electric-dipole,
  quadrupole, and magnetic-dipole susceptibilities and shielding factors for
  closed-shell ions of the He, Ne, Ar, Ni (Cu$^+$), Kr, Pb, and Xe
  isoelectronic sequences}}, At. Data Nucl. Data Tables 28 (1983) 333--340.

\bibitem{Stone.2005}
N.~J. Stone, \href{https://doi.org/10.1016/j.adt.2005.04.001}{{Table of nuclear
  magnetic dipole and electric quadrupole moments}}, At. Data Nucl. Data Tables
  90 (2005) 75--176.

\bibitem{Stone.2014}
N.~J. Stone,
  \href{http://inis.iaea.org/search/search.aspx?orig_q=RN:45029196}{{Table of
  Nuclear Magnetic Dipole and Electric Quadrupole Moments}}, Tech. rep.,
  International Atomic Energy Agency (IAEA), {Nuclear Physics and Radiation
  Physics, INDC(NDS)--0658} (2014).

\bibitem{Antusek.2005}
A.~Antu{\v{s}}ek, K.~Jackowski, M.~Jaszu{\'n}ski, W.~Makulski, M.~Wilczek,
  \href{https://doi.org/10.1016/j.cplett.2005.06.022}{{Nuclear magnetic dipole
  moments from NMR spectra}}, Chem. Phys. Lett. 411 (2005) 111--116.

\bibitem{Skripnikov.2018}
L.~V. Skripnikov, S.~Schmidt, J.~Ullmann, C.~Geppert, F.~Kraus, B.~Kresse,
  W.~N\"ortersh\"auser, A.~F. Privalov, B.~Scheibe, V.~M. Shabaev, M.~Vogel,
  A.~V. Volotka, \href{https://doi.org/10.1103/PhysRevLett.120.093001}{{New
  Nuclear Magnetic Moment of $^{209}\mathrm{Bi}$: Resolving the Bismuth
  Hyperfine Puzzle}}, Phys. Rev. Lett. 120 (2018) 093001.

\bibitem{Fella.2020}
V.~Fella, L.~V. Skripnikov, W.~N{\"o}rtersh{\"a}user, M.~R. Buchner, H.~L.
  Deubner, F.~Kraus, A.~F. Privalov, V.~M. Shabaev, M.~Vogel,
  \href{https://doi.org/10.1103/PhysRevResearch.2.013368}{{Magnetic moment of
  $^{207}$Pb and the hyperfine splitting of $^{207}$Pb$^{81+}$}}, Phys. Rev.
  Research 2 (2020) 013368.

\bibitem{Brown.2014}
B.~A. Brown, W.~D.~M. Rae,
  \href{https://doi.org/10.1016/j.nds.2014.07.022}{{The Shell-Model Code
  NuShellX@MSU}}, Nucl. Data Sheets 120 (2014) 115--118.

\bibitem{KSHELL}
N.~Shimizu, T.~Mizusaki, Y.~Utsuno, Y.~Tsunoda,
  \href{https://doi.org/10.1016/j.cpc.2019.06.011}{{Thick-restart block Lanczos
  method for large-scale shell-model calculations}}, Comput. Phys. Commun. 244
  (2019) 372--384.

\bibitem{Antusek.2020}
A.~Antu{\v{s}}ek, M.~Repisky, \href{https://doi.org/10.1039/D0CP00115E}{{NMR
  absolute shielding scales and nuclear magnetic dipole moments of transition
  metal nuclei}}, Phys. Chem. Chem. Phys. 22 (2020) 7065--7076.

\bibitem{Krane.1976}
K.~S. Krane, S.~S. Rosenblum, W.~A. Steyert,
  \href{https://doi.org/10.1103/PhysRevC.14.650}{{Nuclear magnetic moment of
  $^{65}\mathrm{Ni}$}}, Phys. Rev. C 14 (1976) 650--652.

\bibitem{Rikovska.2000}
J.~Rikovska, T.~Giles, N.~J. Stone, K.~van Esbroeck, G.~White, A.~W\"ohr,
  M.~Veskovic, I.~S. Towner, P.~F. Mantica, J.~I. Prisciandaro, D.~J.
  Morrissey, V.~N. Fedoseyev, V.~I. Mishin, U.~K\"oster, W.~B. Walters, {and
  the NICOLE and ISOLDE Collaboration},
  \href{https://doi.org/10.1103/PhysRevLett.85.1392}{{First On-Line Beta-NMR on
  Oriented Nuclei: Magnetic Dipole Moments of the
  $(\mathit{\ensuremath{\nu}}{\mathit{p}}_{1/2}{)}^{\ensuremath{-}1}$
  $1/{2}^{\ensuremath{-}}$ Ground State in $^{67}$Ni and
  $(\mathit{\ensuremath{\pi}}{\mathit{p}}_{3/2}{)}^{+1}$
  $3/{2}^{\ensuremath{-}}$ Ground State in $^{69}$Cu}}, Phys. Rev. Lett. 85
  (2000) 1392--1395.

\bibitem{Drain.1964}
L.~E. Drain, \href{https://doi.org/10.1016/0031-9163(64)90633-X}{{The magnetic
  moment of $^{61}$Ni}}, Phys. Lett. 11 (1964) 114--115.

\bibitem{Harris.2001}
R.~K. Harris, E.~D. Becker, S.~M. {Cabral de Menezes}, R.~Goodfellow,
  P.~Granger, \href{https://doi.org/10.1351/pac200173111795}{{NMR nomenclature.
  Nuclear spin properties and conventions for chemical shifts (IUPAC
  Recommendations 2001)}}, Pure Appl. Chem. 73 (2001) 1795--1818.

\bibitem{Pyykko.2018}
P.~Pyykk{\"o}, \href{https://doi.org/10.1080/00268976.2018.1426131}{{Year-2017
  nuclear quadrupole moments}}, Molec. Phys. 116 (2018) 1328--1338.

\bibitem{Sternheimer.1972}
R.~M. Sternheimer, \href{https://doi.org/10.1103/PhysRevA.6.1702}{Quadrupole
  shielding and antishielding factors for several atomic ground states}, Phys.
  Rev. A 6 (1972) 1702--1709.

\bibitem{Honma.2009}
M.~Honma, T.~Otsuka, T.~Mizusaki, M.~Hjorth-Jensen,
  \href{https://doi.org/10.1103/PhysRevC.80.064323}{{New effective interaction
  for ${f}_{5}{\mathit{pg}}_{9}$-shell nuclei}}, Phys. Rev. C 80 (2009) 064323.

\bibitem{Mukhopadhyay.2017}
S.~Mukhopadhyay, B.~P. Crider, B.~A. Brown, S.~F. Ashley, A.~Chakraborty,
  A.~Kumar, M.~T. McEllistrem, E.~E. Peters, F.~M. Prados-Est\'evez, S.~W.
  Yates, \href{https://doi.org/10.1103/PhysRevC.95.014327}{{Nuclear structure
  of $^{76}\mathrm{Ge}$ from inelastic neutron scattering measurements and
  shell model calculations}}, Phys. Rev. C 95 (2017) 014327.

\bibitem{Honma.2005}
M.~Honma, T.~Otsuka, B.~A. Brown, T.~Mizusaki,
  \href{https://doi.org/10.1140/epjad/i2005-06-032-2}{{Shell-model description
  of neutron-rich $pf$-shell nuclei with a new effective interaction GXPF 1}},
  Eur. J. Phys. A 25 (2005) 499--502.

\bibitem{Vingerhoets.2011}
P.~Vingerhoets, K.~T. Flanagan, J.~Billowes, M.~L. Bissell, K.~Blaum, B.~Cheal,
  M.~de~Rydt, D.~H. Forest, C.~Geppert, M.~Honma, M.~Kowalska, J.~Kr{\"a}mer,
  K.~Kreim, A.~Krieger, R.~Neugart, G.~Neyens, W.~N{\"o}rtersh{\"a}user,
  J.~Papuga, T.~J. Procter, M.~M. Rajabali, R.~Sanchez, H.~H. Stroke, D.~T.
  Yordanov, \href{https://doi.org/10.1016/j.physletb.2011.07.050}{{Magnetic and
  quadrupole moments of neutron deficient $^{58-62}$Cu isotopes}}, Phys. Lett.
  B 703 (2011) 34--39.

\bibitem{Hergert.2015}
H.~Hergert, S.~K. Bogner, T.~D. Morris, A.~Schwenk, K.~Tsukiyama,
  \href{https://doi.org/10.1016/j.physrep.2015.12.007}{{The In-Medium
  Similarity Renormalization Group: A Novel Ab Initio Method for Nuclei}},
  Phys. Rep. 621 (2016) 165.

\bibitem{Stroberg2019}
S.~R. Stroberg, H.~Hergert, S.~K. Bogner, J.~D. Holt,
  \href{https://doi.org/10.1146/annurev-nucl-101917-021120}{{Nonempirical
  Interactions for the Nuclear Shell Model: An Update}}, Annu. Rev. Nucl. Part.
  Sci. 69 (2019) 307--362.

\bibitem{Hebeler2011}
K.~Hebeler, S.~K. Bogner, R.~J. Furnstahl, A.~Nogga, A.~Schwenk,
  \href{https://doi.org/10.1103/PhysRevC.83.031301}{{Improved nuclear matter
  calculations from chiral low-momentum interactions}}, Phys. Rev. C 83 (2011)
  031301.

\bibitem{Soma2020}
V.~Som{\`{a}}, P.~Navr{\'{a}}til, F.~Raimondi, C.~Barbieri, T.~Duguet,
  \href{https://doi.org/10.1103/PhysRevC.101.014318}{{Novel chiral Hamiltonian
  and observables in light and medium-mass nuclei}}, Phys. Rev. C 101 (2020)
  014318.

\bibitem{Jiang2020}
W.~G. Jiang, A.~Ekstr{\"{o}}m, C.~Forss{\'{e}}n, G.~Hagen, G.~R. Jansen,
  T.~Papenbrock, \href{https://doi.org/10.1103/PhysRevC.102.054301}{{Accurate
  bulk properties of nuclei from $A=2$ to $\infty$ from potentials with
  $\Delta$ isobars}}, Phys. Rev. C 102 (2020) 054301.

\bibitem{Entem.2003}
D.~R. Entem, R.~Machleidt,
  \href{https://doi.org/10.1103/PhysRevC.68.041001}{{Accurate charge-dependent
  nucleon-nucleon potential at fourth order of chiral perturbation theory}},
  Phys. Rev. C 68 (2003) 041001.

\bibitem{Pastore2009}
S.~Pastore, L.~Girlanda, R.~Schiavilla, M.~Viviani, R.~B. Wiringa,
  \href{https://doi.org/10.1103/PhysRevC.80.034004}{{Electromagnetic currents
  and magnetic moments in chiral effective field theory
  ($\chi\,\mathrm{EFT}$)}}, Phys. Rev. C 80 (2009) 034004.

\bibitem{Krebs2019}
H.~Krebs, E.~Epelbaum, U.-G. Mei{\ss}ner,
  \href{https://doi.org/10.1007/s00601-019-1500-5}{{Nuclear Electromagnetic
  Currents to Fourth Order in Chiral Effective Field Theory}}, Few-Body Syst.
  60 (2019) 31.

\bibitem{Seutin2023}
R.~Seutin, O.~J. Hernandez, T.~Miyagi, S.~Bacca, K.~Hebeler, S.~K\"onig,
  A.~Schwenk, \href{https://doi.org/10.1103/PhysRevC.108.054005}{{Magnetic
  dipole operator from chiral effective field theory for many-body expansion
  methods}}, Phys. Rev. C 108 (2023) 054005.

\bibitem{Miyagi2022}
T.~Miyagi, S.~R. Stroberg, P.~Navr{\'{a}}til, K.~Hebeler, J.~D. Holt,
  \href{https://doi.org/10.1103/PhysRevC.105.014302}{{Converged \textit{ab
  initio} calculations of heavy nuclei}}, Phys. Rev. C 105 (2022) 014302.

\bibitem{Stroberg2017}
S.~R. Stroberg, A.~Calci, H.~Hergert, J.~D. Holt, S.~K. Bogner, R.~Roth,
  A.~Schwenk,
  \href{https://doi.org/10.1103/PhysRevLett.118.032502}{{Nucleus-Dependent
  Valence-Space Approach to Nuclear Structure}}, Phys. Rev. Lett. 118 (2017)
  032502.

\bibitem{Morris2015}
T.~D. Morris, N.~M. Parzuchowski, S.~K. Bogner,
  \href{https://doi.org/10.1103/PhysRevC.92.034331}{{Magnus expansion and
  in-medium similarity renormalization group}}, Phys. Rev. C 92 (2015) 034331.

\bibitem{Parzuchowski2017}
N.~M. Parzuchowski, S.~R. Stroberg, P.~Navr{\'{a}}til, H.~Hergert, S.~K.
  Bogner, \href{https://doi.org/10.1103/PhysRevC.96.034324}{{Ab initio
  electromagnetic observables with the in-medium similarity renormalization
  group}}, Phys. Rev. C 96 (2017) 034324.

\bibitem{Pritychenko.2012}
B.~Pritychenko, J.~Choquette, M.~Horoi, B.~Karamy, B.~Singh,
  \href{https://doi.org/10.1016/j.adt.2012.06.004}{{An update of the $B$(E2)
  evaluation for $0_1^+ \rightarrow 2_1^+$ transitions in even–even nuclei
  near $N\sim Z\sim 28$}}, At. Data Nucl. Data Tables 98 (2012) 798--811.

\bibitem{Goeppert-Mayer.1955}
M.~Goeppert-Mayer, J.~Jensen, {Elementary Theory of Nuclear Shell Structure},
  John Wiley \& Sons, Inc., New York, 1955.

\bibitem{NuHamil}
T.~Miyagi, \href{https://doi.org/10.1140/epja/s10050-023-01039-y}{{NuHamil: A
  numerical code to generate nuclear two- and three-body matrix elements from
  chiral effective field theory}}, Eur. J. Phys. A 59 (2023) 150.

\bibitem{imsrg++}
S.~R. Stroberg, \href{https://github.com/ragnarstroberg/imsrg}{{IMSRG++}}
  (2018).

\end{thebibliography}

\end{document}